# Development and Commissioning of a Compact Cosmic Ray Muon Imaging Prototype

Xujia Luo[a], Quanxiao Wang[a], Kemian Qin[a], Heng Tian[a], Zhiqiang Fu[a], Yanwei Zhao[a], Zhongtao Shen[c], Hao Liu[c], Yuanyong Fu[d,e], Guorui Liu[a], Kaiqiang Yao[a], Xiangping Qian[a], Jian Rong[a,b], Weixiong Zhang[f], Xiaogang Luo[f], Chunxian Liu[f], Xiangsheng Tian[f], Minghai Yu[g], Feng Wu[g], Jingjing Chen[h], Juntao Liu[a,b,*], Zhiyi Liu[a,b,*]

[a]*School of Nuclear Science and Technology, Lanzhou University, Lanzhou, Gansu, China, 730000*
[b]*Frontiers Science Center for Rare Isotopes, Lanzhou University, Lanzhou, Gansu, China, 730000*
[c]*State Key Laboratory of Particle Detection and Electronics, University of Science and Technology of China, Hefei, Anhui, China,  230026*
[d]*China Institute of Atomic Energy, Beijing, China, 102413*
[e]*Beijing Isotope Nuclear Electronic Machine Co.Ltd, Beijing, China, 102412*
[f]*Third Institute Geological and Mineral Exploration of Gansu Provincial Bureau of Geology and Mineral Resources, Lanzhou, Gansu, China, 730050*
[g]*Hezuo Zaozigou Gold Mine Co.,Ltd. of Gansu Province,Gannan, Gansu, 747099*
[h]*Gansu Xibei Gold Co.,Ltd, Lanzhou, Gansu, 730300*

**Abstract**

Due to the muon tomography's capability of imaging high $Z$ materials, some potential applications have been reported on inspecting smuggled nuclear materials in customs. A compact Cosmic Ray Muons (CRM) imaging prototype, Lanzhou University Muon Imaging System (LUMIS), is comprehensively introduced in this paper including the structure design, assembly, data acquisition and analysis, detector performance test, and material imaging commissioning etc. Casted triangular prism plastic scintillators (PS) were coupled with Si-PMs for sensitive detector components in system. LUMIS's experimental results show that the detection efficiency of an individual detector layer is about 98%, the position resolution for vertical incident muons is 2.5 mm and the angle resolution is 8.73 mrad given a separation distance of 40.5 cm. Moreover, the image reconstruction software was developed based on the Point of Closest Approach (PoCA) to detect lead bricks as our target. The reconstructed images indicate that the profile of the lead bricks in the image is highly consistent with

---

*Corresponding author:

*Email addresses:* zhiyil@lzu.edu.cn (Zhiyi Liu), ljt@lzu.edu.cn (Juntao Liu)

the target. Subsequently, the capability of LUMIS to distinguish different materials, such as Pb, Cu, Fe, and Al, was investigated as well. The lower limit of response time for rapidly alarming high-$Z$ materials is also given and discussed. The successful development and commissioning of the LUMIS prototype have provided a new solution option in technology and craftsmanship for developing compact CRM imaging systems that can be used in many applications.

## 1. Introduction

Cosmic Ray Muons (CRM) are the decay product of the pions which are generated by the interaction of the interstellar-space high-energy protons with earth's atmosphere. The flux of CRM at sea level is approximately 10,000 m$^{-2}$min$^{-1}$ with average energy at 4 GeV. CRM are also the most numerous particle reaching sea level among all the secondary charged particles of cosmic rays [1]. There are some factors, such as the latitudes on earth [2], solar activities [2][4][5] and temperature [6][7][8], affecting the CRM flux. CRM have relatively strong penetrability in rocks or metals because of its high energy, broad energy range up to TeVs and small interaction cross section with matters [9]. As an abundant natural source, CRM have been attracting researchers' attention due to its extraordinary advantages comparing with traditional non-destructive imaging technology, such as X-ray [10][11][12][13]. The first application of CRM was performed in 1955 by George *et al.* [14] via using Geiger counter to surmise the thickness of the rock overburden above a tunnel of a mineral mountain located in Australia. In 1970, Alvarez *et al.* scanned the second largest pyramid via reconstructing CRM tracks by using a spark chamber and firstly illustrated the advantage of muon application in the field of archaeology [15]. Since 1995, the application of CRM imaging technology had been introduced to geophysics by imaging the geological structure of a volcano by Nagamine *et al.* [16]. In 2003, two segmented detectors made of plastic scintillators were used to detect the volcano, Mt. Asama, which team proved that volume occupancy in the region of a crater is less than 30% [17]. A hidden chamber in the Khufu pyramid was discovered via comparing experimental data with the Monte-Carlo simulation result [18]. The shape of bedrock underneath the Eiger glacier was reconstructed by Nishiyama *et al.* using emulsion film detectors [19]. These applications are all based on the principle of absorption of muons crossing targets.

CRM also have significant applications at border security and defense since the muon tomography based on Multiple Coulomb Scattering (MSC) property was first introduced by Borozdin in 2003 [20]. In 2011, the Gas Electron Multiplier (GEM) detectors were built to image medium-$Z$ and high-$Z$ targets by Kondo *et al* [21]. In 2014, a team from Tsinghua University produced a CRM imaging prototype based on Multi-gap Resistive Plate Chamber (MRPC) [22] and conducted three-dimensional imaging experiments for customs inspection. However, it may not be convenient to employ and maintain in some scenarios because of some limits of gas detectors [11]. Canadian scientists developed their muon tomography system, Cosmic-Ray Inspection and Passive Tomography (CRIPT) [23], using extruded plastic scintillators (PS) with wavelength shifting (WLS) fibers. In the CRIPT system, CRM momentum are also measured by its spectrometer layers, to effectively reduce the response time for its heavy material alarm function. Mahon's team built a muon prototype based on scintillating fiber, they completed imaging of stainless steel, lead and uranium [24]. A 320 mm×320 mm×25 mm plastic scintillator panel with fibers in grooves was built as the muon detector by a team from Oregon State University [25], this detector achieved 1 cm position resolution. Liang *et al*. [26] built an experimental setup to compare and analyze the position resolution of three different PS structures (triangular prism, cuboid and flat plate) with fiber assembled in groove.

As seen above, although researches on muon imaging detection systems and their applications have been conducted extensively and in-depth, there are still some outstanding problems that need to be paid attention to and improve for the systems based on scintillator. For example, the CRIPT solution [23] is a workable solution in terms of low-cost construction and decent performance, however, its sensitive component, extruded PS triangle bars, is only supplied by very few institutes and is not available currently to procure for researchers in some countries. In addition, in the CRIPT system, scintillating light is collected and guided via WLS fibers coupled with multi-pixel PMT units, extreme caution should always be given in handling, assembling and transporting because the WLS fiber is fragile. The WLS fiber also requires extra space in the light-proof chamber due to the requirement on minimal bending radius. From the cost aspect, such a solution may not have significant advantage in producing small quantities over other solutions. Therefore, we decided to build a brand-new, compact muon imaging detector prototype with lower cost, LUMIS. LUMIS targets at achieving the following primary objectives:

1. using casted PS directly coupled with Si-PM chips as sensitive detector components in the system;

2. based on such a prototype, studying its performance and evaluating whether it can meet the imaging requirements, such as in customs application scenarios;

3. developing a mature, low-cost, highly integrated, and stable electronic data acquisition system since no turn-key solution for a muon imaging system is commercially available so far;

4. thoroughly considering all kinds of application scenarios, designing the system meeting the requirements of easy-assembly and -disassembly, easy-to-operate, maintenance free, and vibration/shock resistance etc.

This paper comprehensively introduces the development and commissioning of LUMIS in detail as a practical solution of building a low-cost, compact and robust detector different from existing detector systems such as Ref. [23]. Section 2 of this paper describes the hardware and electronics design of the LUMIS. Section 3 explains the reconstruction approach of muon tracks. Section 4 summarizes LUMIS' performance such as muon detection efficiency, position resolution and angular resolution. Section 5 presents the Monte Carlo simulation setup and results. Section 6 illustrates the results of experimental imaging, material identification and high $Z$ material rapid alarm.

## 2. Detection system
### 2.1 Detector hardware composition

We selected SP-101 plastic scintillator [27], as sensitive material. The light yield of this type scintillator is about 10 000 ph MeV$^{-1}$, and the light attenuation length is estimated as about 100 cm for our slim triangle bars. The bulk plastic scintillators were cut and milled to the triangular prisms with the base face of an isosceles right triangle as shown in Figure 1. The triangle is 30 mm in base line, 15 mm in height, and 480 mm in total length. All the surfaces of the triangular prism scintillators (TPS) were polished and applied with transparent silicone oil specially for optical coupling purpose. The lateral faces and one of the base face of TPS were covered with Enhanced Specular Reflector (ESR) film while another base face was coupled with Si-PM by optical coupling grease. This bar unit was used as a basic detector channel. 32 channels were laid out to a plane detector with a sensitive area of 480 mm×480 mm as shown in Figure 2. All the plastic scintillating materials with their Si-PMs and auxiliary circuit boards were encapsulated in a light-proof aluminum case. Such two detector planes are orthogonally stacked together as shown in Figure 2 to form a "super layer" providing a two- dimensional muon hitting position.

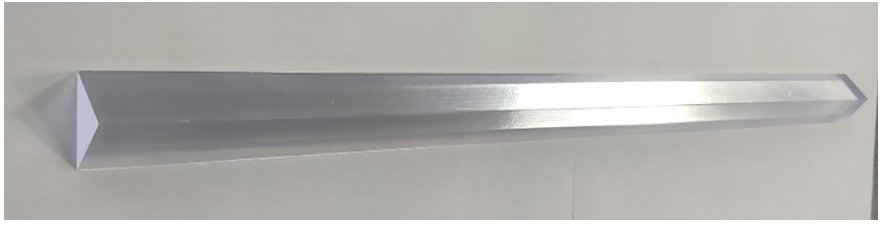

Figure 1: Physical diagram of a TPS

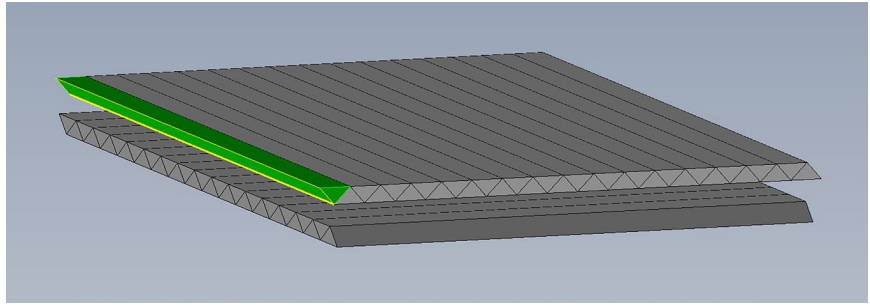

Figure 2: Schematics of TPSs splicing

2.1.1 Geometry

In our detector system, four super layers were stacked parallel at certain spatial distances supported by 4 threaded rods perpendicular to the ground as shown in Figure 3 and Figure 4. The upper two super layers with a separation distance of 40.5 cm is used to measure incident muon tracks to form a "Upper Tracker (UT)". The lower two super layers with the same distance is called "Lower Tracker (LT)" to measure the tracks coming out from a target area. The space between "UT" and "LT" is "Target Area". Targets, such as the lead bricks, are placed here. A layer of 5 cm-thick lead bricks is laid below the bottom detector plane and above the ground to shield gamma ray background from the ground.

**2.2 Detector electronics and data acquisition**

2.2.1 General description

A Sensl 60035 Si-PM chip [28] is coupled with one base face of a basic sensitive channel via optical coupling grease. The active area of this Si-PM is 6 mm×6 mm, the Photon Detection Efficiency (PDE) is about 41% at 420 nm, the dark count rate is about 1200 kHz, and the gain at anode to cathode readout mode is $3 \times 10^6$. The Si-PM chip is used to read out scintillating light and perform the photon-electron conversion. As mentioned above, another base face is covered by a piece of ESR film to avoid the scintillating light from escaping. All the Si-PM chips are soldered on one side of a rectangular PCB board, while their auxiliary circuits, such as filters, are distributed on another side of the board. Such a design is for easily overall installation and commissioning (Si-PM electronics are shown in Figure 5 and the back of the PCB board is soldered with Si-PMs).

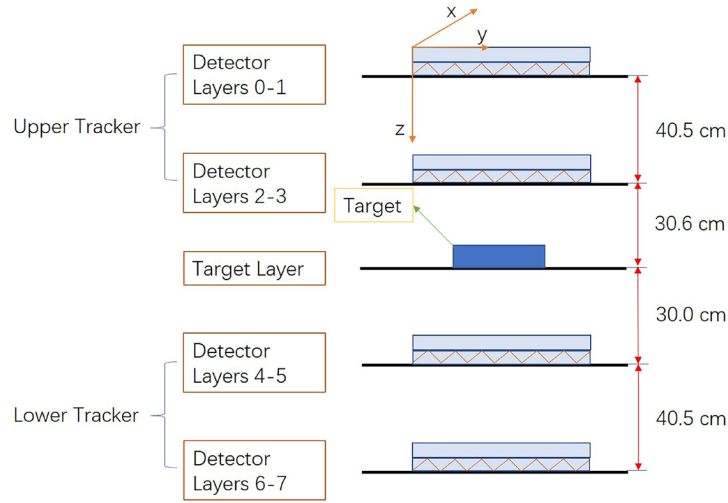

Figure 3: Schematics of the detectors placement

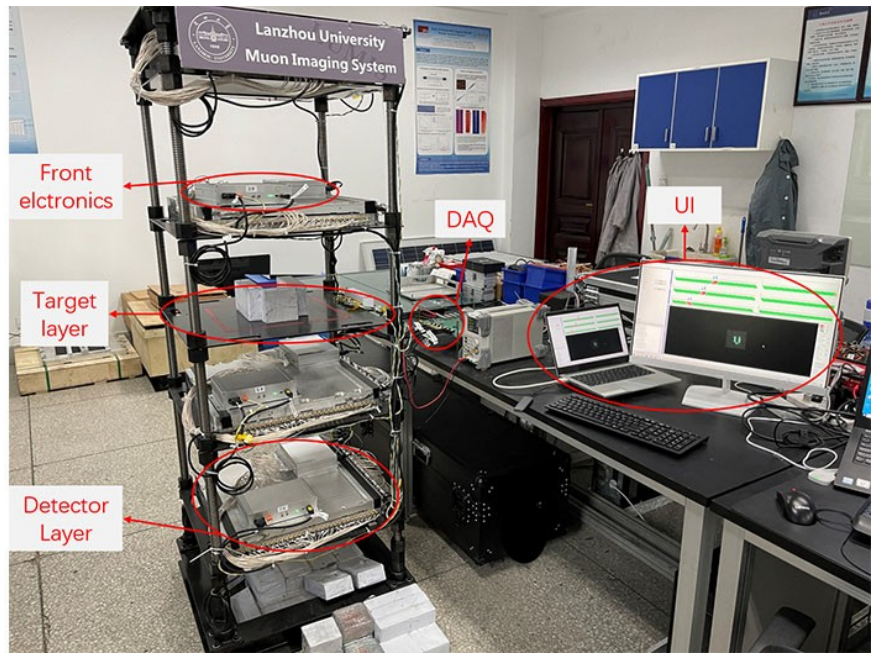

Figure 4: The photograph of LUMIS.

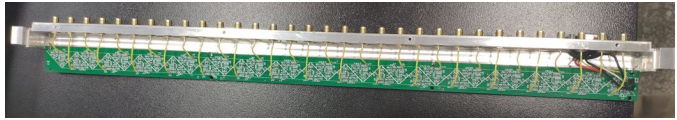

Figure 5: Si-PM electronics, the back side of the PCB is soldered with Si-PMs

The data acquisition system (DAQ) is developed by a team from the University of Science and Technology of China [29]. DAQ has 8 data acquisition boards (also called Front-End Electronics, shown in Figure 6a) and a master board (shown in Figure 6b). Each data acquisition board has 36 electronic channels and processes signals from one detector layer (with 32 channels connected to the TPSs bars and 4 channels spared), the application specific inte- grated circuit (ASIC), SPIROC, is used for processing the 36-channel signals. The charge and time signals are stored in 16 depth switched capacitor arrays (SCAs) respectively, and then digitized by an embedded 12-bit Wilkinson Analog-to-Digital Converter (ADC). The data acquisition board processes the signals from the Si-PM chips which exceed the preset threshold. The master board collects data from the 8 data acquisition boards, performs coincidence and then transfers data to a personal computer. The overall structure schematics of the detector system is shown in Figure 7. To achieve a spatial resolution of 2.5 mm, the energy resolution of the readout hierarchy should be better than 0.55 MeV, the energy resolution of DAQ is 0.14 MeV [29]. More features and details about LUMIS's DAQ can be found in Ref. [29].

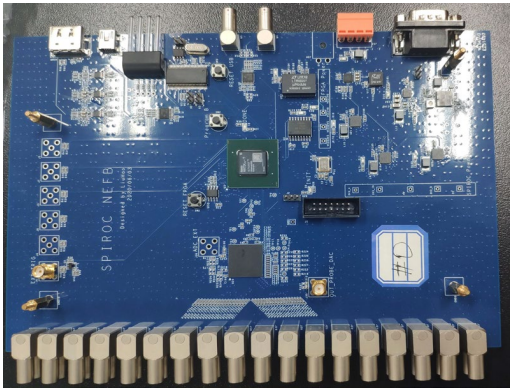
(a)

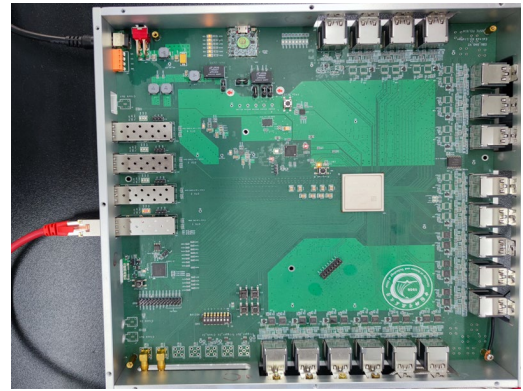
(b)

Figure 6: (a) Data acquisition board (Front-End Electronics) and (b) the master board of DAQ

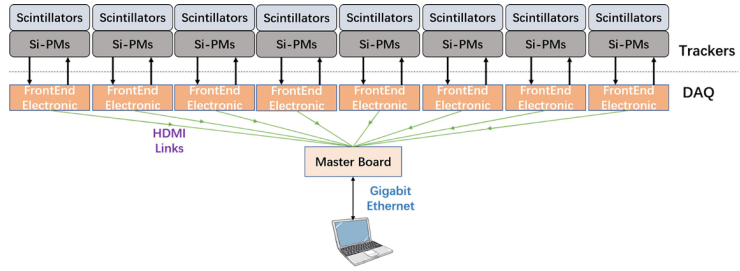

Figure 7: Diagram of detector system connection

2.1.2 Software for DAQ

The operating software for LUMIS'DAQ programms in Python combined with a graphical user interface (GUI) based on PyQt5 (version=5.15) [30]. The LUMIS platform has 3 primary modules. (1) Communication module: it communicates with the electronic system through Ethernet. This module mainly realizes 3 sub-functions: (a) generating the parameters for configuring the data acquisition boards; (b) sending instructing commands and receiving data via TCP/IP protocol; (c) decoding the received binary data. (2) The second module is the data storage and processing module, including high-efficiency data storage solutions, data processing algorithms, reconstruction algorithms for the target objects, and fast-response alarm algorithms for high $Z$ materials. (3) The third module is the GUI module which implements the functions of GUI display, event visualization, and 3D structure drawing of the target object. Figure 8 shows the software interface screen snapshot.

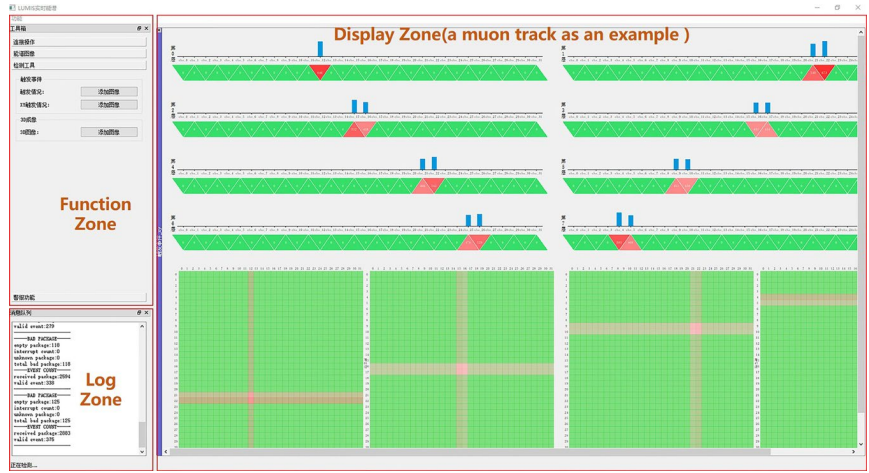

Figure 8: DAQ Graphical User Interface (GUI) with the visualization of a muon event hit

**2.3 Event trigger modes**

Since each electronic channel is unavoidably accompanied by noises, it is necessary to set an appropriate trigger threshold for every channel. Only the signals that exceed their thresholds will be collected, processed and sent to the master board. Based on our test, when the threshold was set as 16 pC, we detect as many good muon events as possible while keeping relatively low dead time. The master board performs events coincidence treatment for signals from 8 data acquisition boards according to the preset coincidence logic [29]. In the imaging collection mode of this system, Layers 0, 3 and 7 were chosen to perform coincidence measurement while the rest layers were set as the self-trigger mode. Only the 3 layers are selected to trigger an event candidate because we need to estimate and derive detector efficiency (see Section 4.1) and also use the passing rate of good events as an evaluation on DAQ's health status. A valid muon event will be recognized and recorded when signals are detected in Layers 0, 3, and 7 during the same 4 μs time window. Thus good CRM events are selected to the greatest extent, while the interference of other radiation sources, such as natural background gamma rays and electronic noise can be reduced.

## 3. Data analysis
### 3.1 Position reconstruction

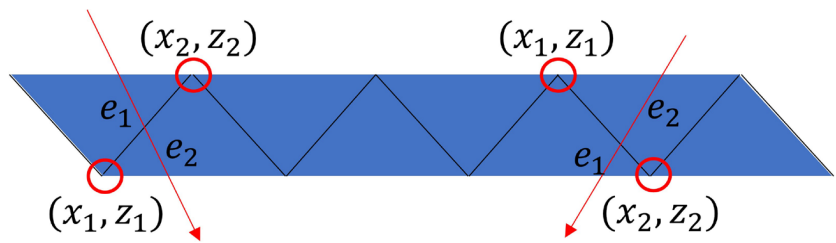

Figure 9: Diagram of muon hit position calculation

This section introduces how a hit position is measured in the position sensitive detectors. Because the energy deposited by a muon in scintillator is proportional to the penetration length in the scintillator and the collected light signal amplitude is proportional to the path length, the position of the muon hits in the scintillator can be calculated with the assistance of this relationship and the adjacent structure of the TPSs. To be a sample, assuming that the incident muon zenith angle is less than 45°, as shown in Figure 9, the hit position can be expressed as:

$$x = \frac{x_1 * e_1 + x_2 * e_2}{e_1 + e_2} \quad (1)$$

$$z = \frac{z_1 * e_1 + z_2 * e_2}{e_1 + e_2} \quad (2)$$

where $(x_1, z_1)$ and $(x_2, z_2)$ are the coordinates of the vertexes of the TPSs cross-section, $e_1$ and $e_2$ are the energy deposited inside two adjacent scintillators, which is proportional to the optical signal amplitude.

## 3.2 Experimental spectrum

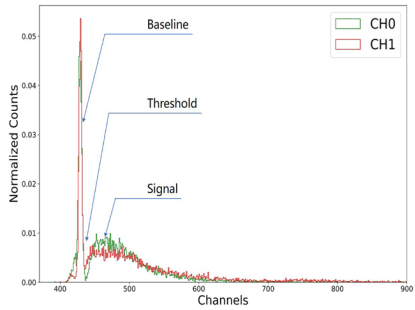

(a)

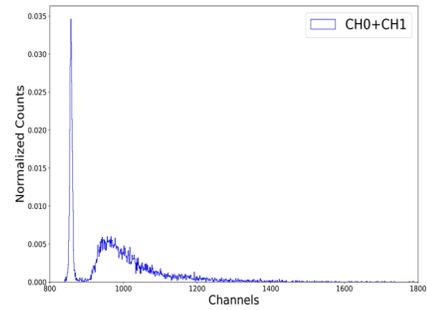

(b)

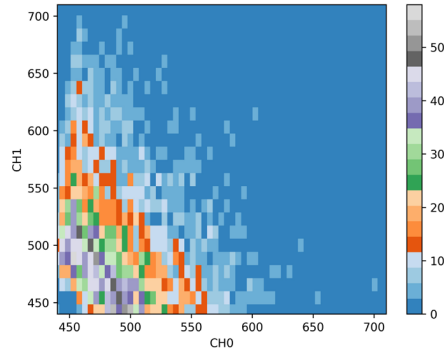

(c)

Figure 10: (a) Spectra of two adjacent TPSs coincidence, (b) sum of two TPSs spectrum, (c) correlation of two TPSs

Figure 10 shows signal amplitude in each bar and how correlated the neighbor bars are in coincidence. Figure 10a is the signal peak height (PH) spectra of the two adjacent channels. In this figure, the peak on the left refers to the baseline, the platform on the right of the peak is a result of the muon energy deposition. The neck between peak and platform in Figure 10a is the threshold to select muon events, only the signal amplitude values larger than this threshold could be muon events candidates. Figure 10b illustrates the spectrum of the signal PH sum from the two adjacent channels. The peak on the left is the sum of two baselines and the peak on the right results from the sum of the energy of muon deposited in the two adjacent TPSs. Figure 10c shows the negative correlation between the two neighbor bars (the baselines of two channels have been removed from the data), which is expected according to the triangle structure of the bars.

### 3.3 Muon track reconstruction

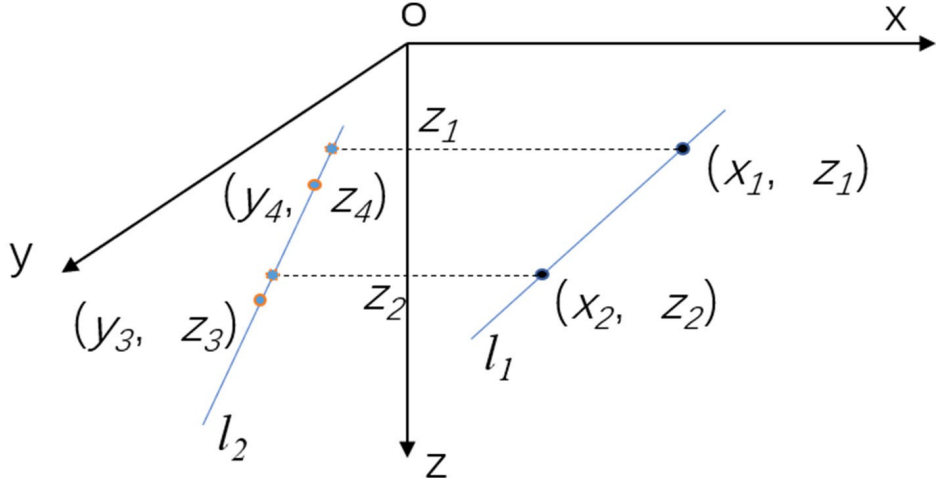

Figure 11: Diagram for entry trajectory calculation

As shown in Figure 11 $l_1$ and $l_2$ are the projected trajectory of one incident muon in X-Z plane and Y-Z plane. They are determined by the 2 X-Z coordinates, $(x_1, z_1)$ and $(x_2, z_2)$, and 2 Y-Z coordinates, $(y_3, z_3)$ and $(y_4, z_4)$, which are hit positions on two super layers of "UT". Based on the function of $l_2$, we calculate the $y_1$ ($y_1 = \frac{(z_1-z_3)(y_4-y_3)}{z_4-z_3} + y_3$) and $y_2$ ($y_2 = \frac{(z_2-z_3)(y_4-y_3)}{z_4-z_3} + y_3$) given $z = z_1$ and $z = z_2$ respectively. Then the two 3-dimension coordinates $(x_1, y_1, z_1)$ and $(x_2, y_2, z_2)$ on the muon incident trajectory and the incident track are calculated. Similarly, the exiting track is gotten by hits on the "LT".

## 4. Detection Performance
### 4.1 Detection Efficiency

The statistic of muon rays is one of the bottlenecks restricting muon imaging application, therefore, it is necessary to study and improve the detection efficiency as much as possible. In this work, a muon event is measured by the coincidence method when Layers 0, 3, 7 are triggered simultaneously. We analyzed about 276,000 muon events and counted the number of fired layers in each event as shown in Figure 12. The number of the events with all eight- layers triggered monopolized as 90.31% of total events. Each layer has only two scenarios, fired or not, which probability depends on intrinsic detection efficiency. Assuming that each layer has the same detection efficiency, the statistics result of fired layers obeys the binomial distribution, then the detection efficiency can be calculated. Since the Layers 0, 3, and 7 are triggered in all events, the probability of the events in which all 8 layers were triggered satisfies the formula:

$$p^5 = 0.9031 \qquad (3)$$

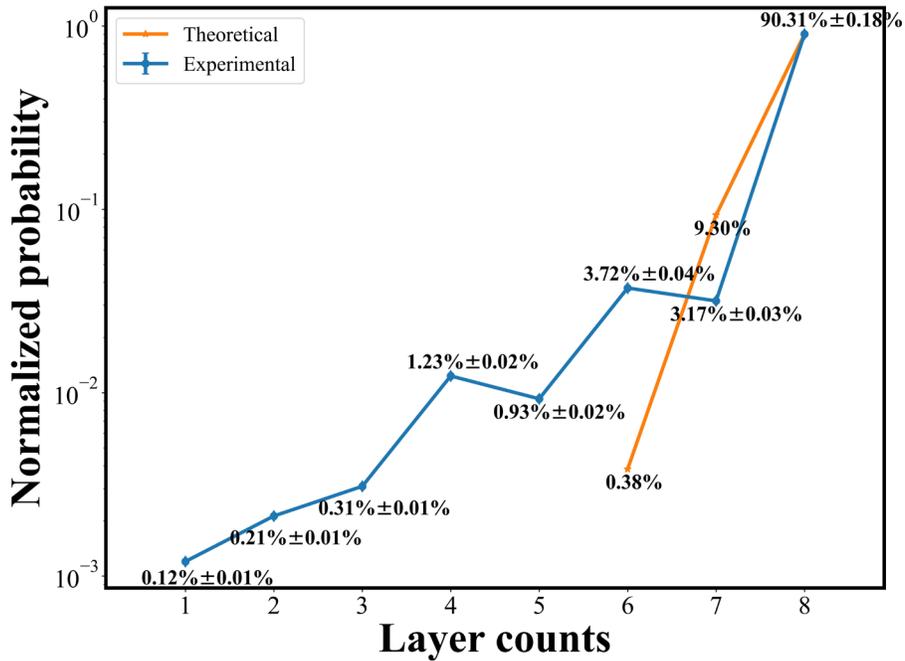

Figure 12: Statistics on the number of trigger events in different layers

Therefore, the detection efficiency $P$ is calculated as 97.98%. The theoretical distribution of other combinations is shown in Figure 12 with orange line (the theoretical probability of being triggered for less than 6 layers is too small to be drawn in the figure). In addition, some events with either one or two fired layers occupied 0.33%, which are caused by the inner-logic of the DAQ system. Based on our study of DAQ, there is one explanation for this phenomenon. Since the coincident time window is set to 4 μs in DAQ, if a muon event arrives right before the end of a previous time window, the time point when master board receives and responds to all data acquisition boards may fall into the next time window. In this scenario, the DAQ will record the signals arriving only during the second time window regardless whether or not they satisfied the preset coincidence condition, thus it leads to the records of less than 3 layers be triggered.

The events to reconstruct good tracks must satisfy the requirements that one layer should be triggered less than or equal to 2 bars, and the 2 bars condition should be 2 neighbor bars. The percentage of the events under this filter criteria is 50.18% of the total number of events we measured, so the detection efficiency for reconstructing muons tracks should be 91.74% ($P^8 = 50.18\%$).

**4.2 Position Resolution**

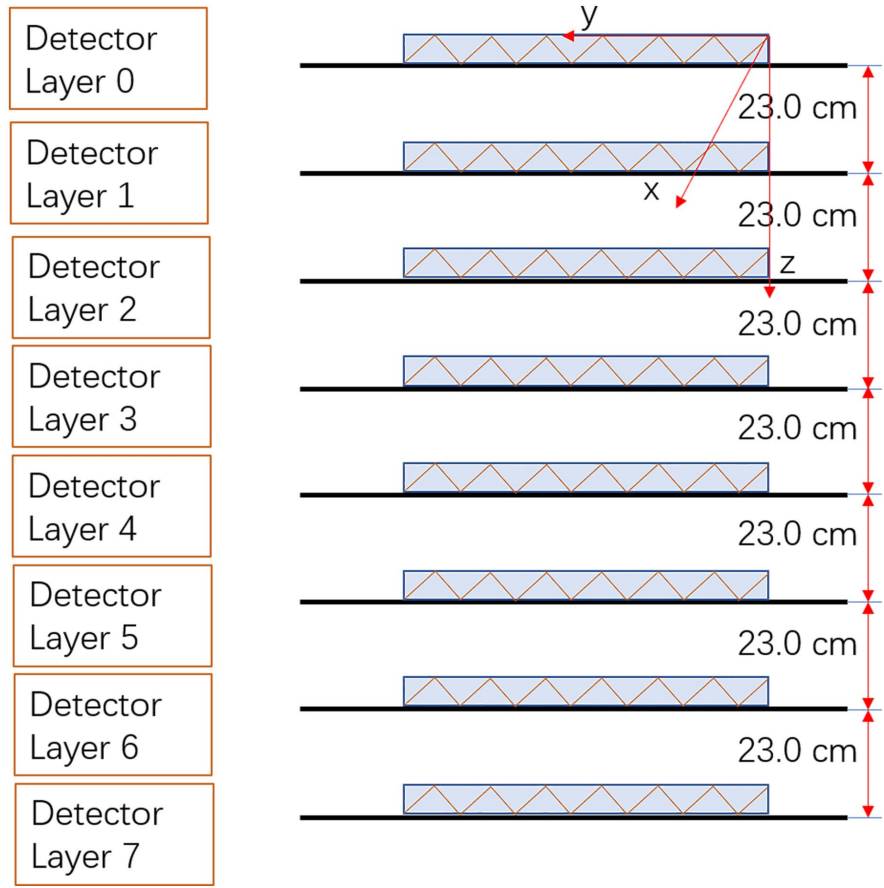

Figure 13: Schematic diagram of 8 layers of detectors placed in the same direction

In the position resolution test, we use Layer 4 as the target layer. Rest seven layers are used to determine muons' trajectories by linear fitting seven hit position coordinates. Since seven positions were used to fit a straight line, we assume that the fitting position of the target layer, $x_{fit}$, is close enough to the true position of the hit. Therefore, the residual distribution is expressed by $x_{fit} - x_{mea}$ where $x_{\text{mea}}$ is the measured position. In order to investigate the position resolution of LUMIS, all eight detector layers are placed parallel in the same direction as shown Figure 13. The residual distribution of LUMIS is shown in Figure 14, the standard deviation, $\sigma$=2.50 mm refers to the LUMIS position resolution.

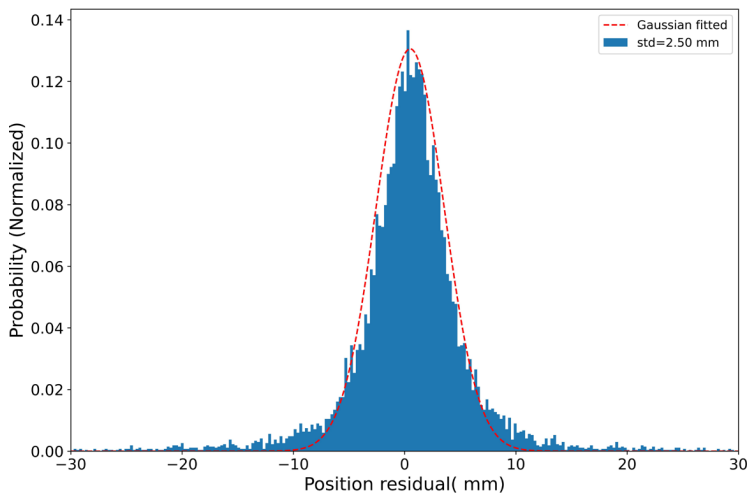

Figure 14: The position residual distribution of LUMIS

**4.3 Angular resolution**

The angular resolution is then derived from the hit position resolution and the separation distance of planes. The distances between 2 super-layers for "UT" and "LT" both are 40.5 cm according to Figure 3. so the angular resolution is 8.73 mrad. For the muons with the energy of 3 GeV/c, scattering angles ($\sigma(\Delta\theta)$) of muons after passing through 10 cm thickness in steel and lead are around 11 mrad and 20 mrad, respectively [20]. Therefore, LUMIS has the ability to distinguish these two materials with a thickness of 10 cm through scattering angles.

**5. Simulations**

A simplified Monte-Carlo (MC) simulation on LUMIS was conducted in the framework of Geant4 to theoretically study its comprehensive performance, understand the difference between an ideal detector and the realistic detector and, further, optimize the detector configuration. In the simulation, we employ the Cosmic-RaY shower generator (CRY) [31] to generate muon rays at sea level above LUMIS.

## 5.1 Geometry model

The basic detector unit used in the simulation was a triangular prism, as the same shape and dimension as the TPS used in LUMIS. 256 basic detector bar units were created and organized to form four super layers same as the real LUMIS. For simplicity with key features kept, aluminum shells of each detector layers and the supporting frame are not considered in software modelling. The deposited energy at each bar was used to calculate the hit position of a muon event rather than counting total scintillated photons to further reduce the computing time consumption. The imaging target was formed via two 20 cm×10 cm×5 cm lead bricks and one 10 cm×10 cm×5 cm lead brick as shape "U" placed in the middle of upper and lower layers.

## 5.2 Simulation Results

The incoming and outgoing muon trajectories were calculated by using energy information to solve Eq. $x=\frac{x_1*e_1+x_2*e_2}{e_1+e_2}$ (1 and $z=\frac{z_1*e_1+z_2*e_2}{e_1+e_2}$ (2. A simple Point of Closet Approach (PoCA) algorithm [32] was employed to reconstruct the target lead bricks. After generating around 75000 muon events, equivalent to approximate 10 hours muon events for LUMIS located in Lanzhou University (1500 meters above sea level) in experiment, the reconstructed imaging result of the target is shown in Figure 15. It is clearly shown the profile in the result perfectly fits the original position of the target, meanwhile, the density of PoCA points reflects the thickness of the target which proved the feasibility of using LUMIS to image the high $Z$ targets.

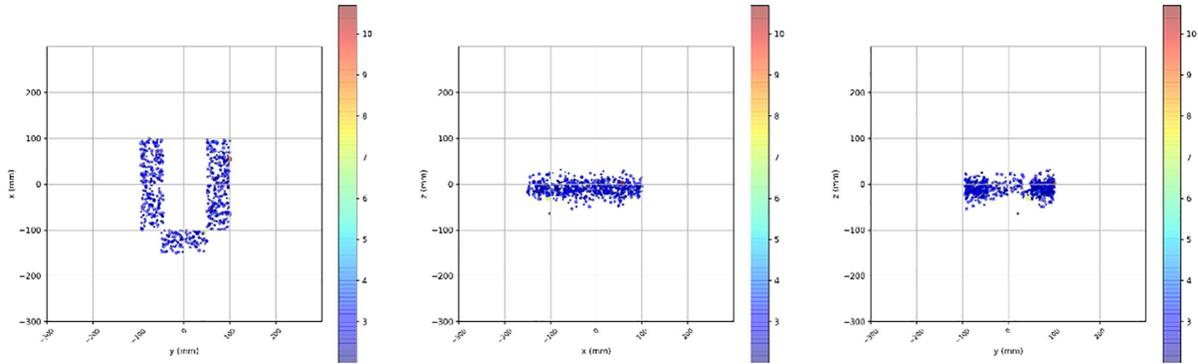

Figure 15: Simulation result of 3D Image reconstruction based on PoCA for ideal detectors in which hit position is derived based on deposited energy in bars

# 6. Experiments
## 6.1 Imaging Experiments

The scattering angles distribution of muons passing through matter is varying for different materials with Z. Its standard deviation is expressed by Eq. $\sigma(\Delta\theta) = \frac{13.6 MeV}{\beta cp}\sqrt{\frac{L}{X_0}}[1 + \ln\frac{x}{X_0\beta^2}]$  ( 4 [33].

$$\sigma(\Delta\theta) = \frac{13.6 MeV}{\beta cp}\sqrt{\frac{L}{X_0}}[1 + \ln\frac{x}{X_0\beta^2}] \quad (4)$$

where $\beta$ is the ratio of muon's velocity over the speed of light, $p$ is muon momentum in MeV/c, $L$ is the path length of muon through the material in g/cm$^2$, and $X_0$ is the radiation length of the material in g/cm$^2$.

Two 20 cm×10 cm×5 cm lead bricks and one 10 cm×10 cm×5 cm lead brick are organized to form a "U" shape as the target for imaging which is as the same as in the simulation as shown in Figure 16. Figure 17 shows the reconstructed image based on 24 hours of data collection on the "U"-shaped lead bricks. Only PoCA points generated by muon events with scattering angles greater than 5 degrees are shown in the figure. It can be seen that in the *X-Y* plane, the shape of the "U" can be clearly identified and the boundary is well defined. Figure 18 shows the projection of PoCA points along *X* and *Y* axes, and the values in *X* and *Y* axes reflect the thickness of lead in two axes. However, the large ambiguity along the *Z*-direction of the image is due to the PoCA points calculation interference of the too-small scattering angle events, making some jointing points fallen outside the target area. The ambiguity can be reduced by two solutions:

- Improve the position resolution of the detector. Figure 15 and Figure 19 show the Monte Carlo simulation results based on an ideal detector and a simulated realistic detector in which counted the measured hit position resolution. The fact that the former barely has ambiguity along *Z* direction while the latter has significant ambiguity demonstrates that the effect of position resolution plays an important role in quality of the reconstructed image.

- Increase the sensitive area of the detector and hence acceptance angles so that muons with larger scattering angles can be detected so that they can define the profile better along the *Z* direction.

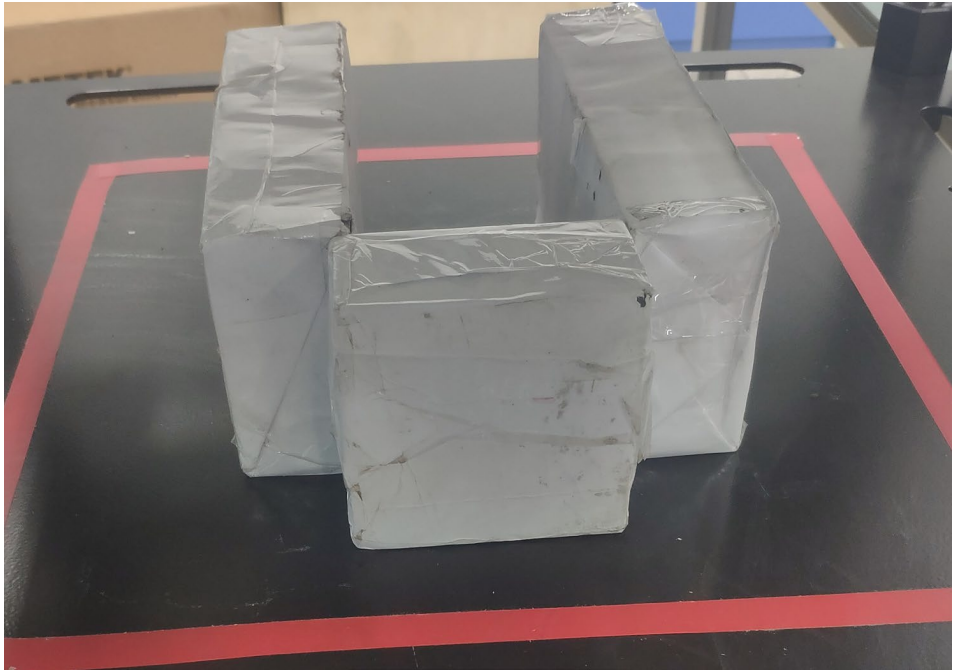
Figure 16: U-shaped lead bricks physical picture

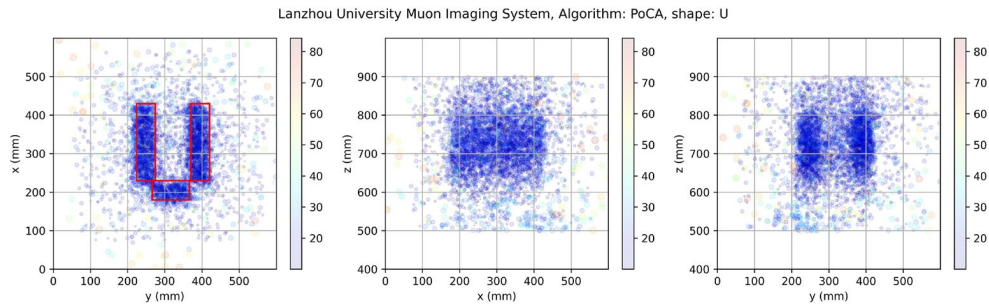
Figure 17: Image reconstruction of U-shaped lead

Furthermore, four bricks made of Pb, Cu, Fe and Al (10 cm×10 cm×5 cm) are employed for verifying the detector's material identification ability, the displacement diagram and reconstruction figures are shown in Figure 20 and Figure 21. It can be seen that the point density of Pb is significantly greater than the other three materials in the square area of themselves, and the point density of Al is significantly smaller than the other three materials. The difference of point density between Cu and Fe is not significant, but it can still be distinguished in Figure 21 with the help of rectangle. So we could identify different materials by the point density in the same measure time.

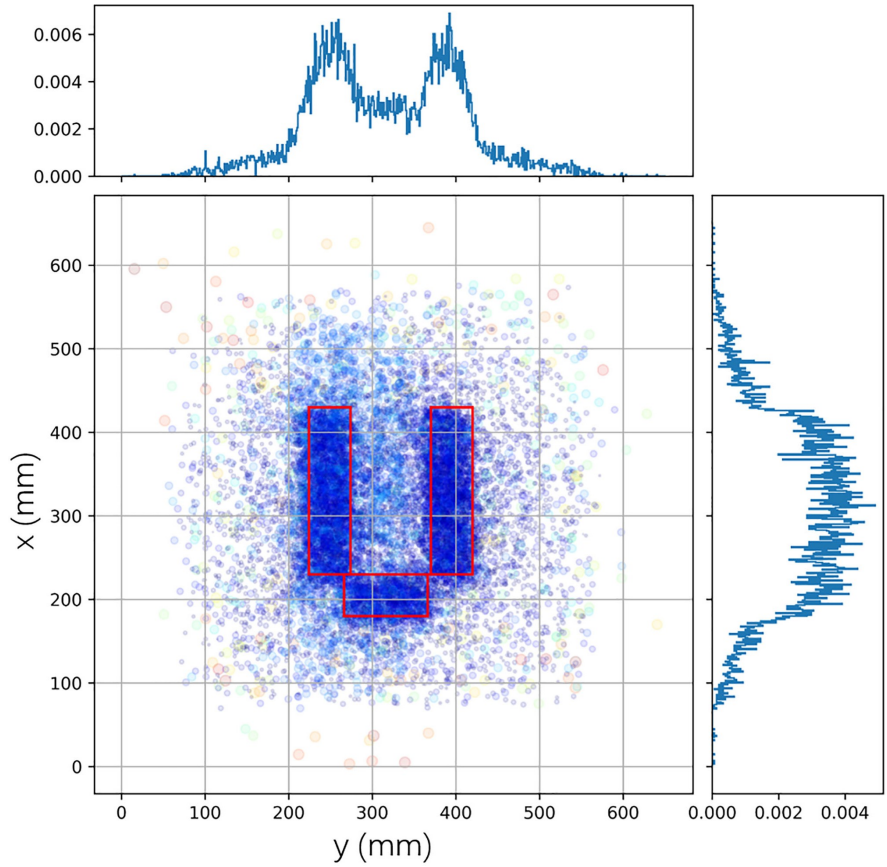

Figure 18: The top view of "U" shape lead with the histograms over X and Y two axes are the distribution of PoCA points reflects the thickness of the material

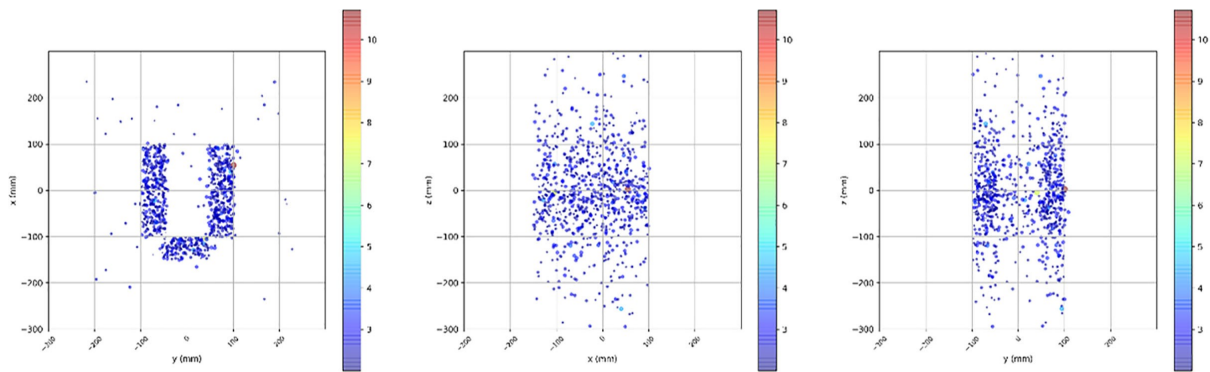

Figure 19: Reconstruction image based on simulation combined with detector effect

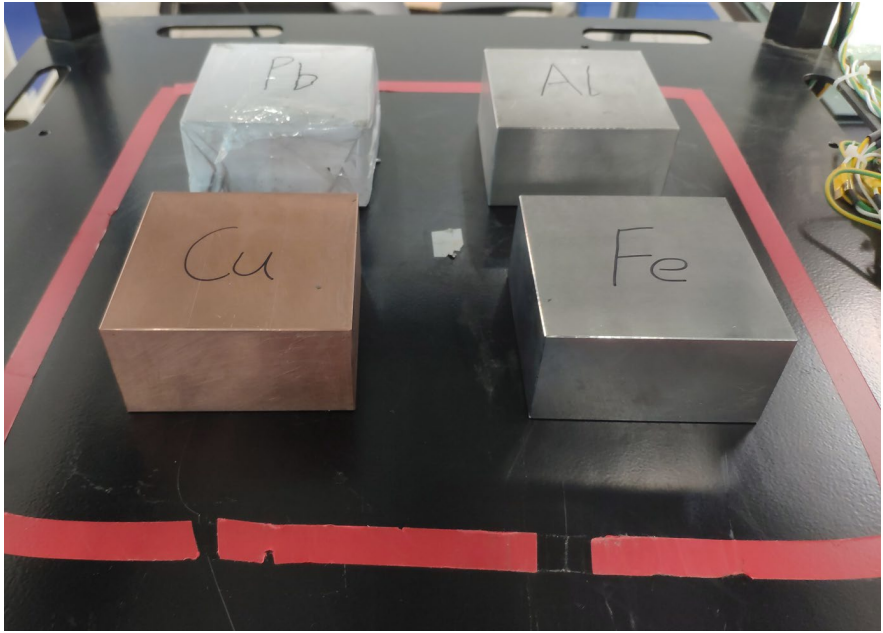

Figure 20: Photograph of different materials

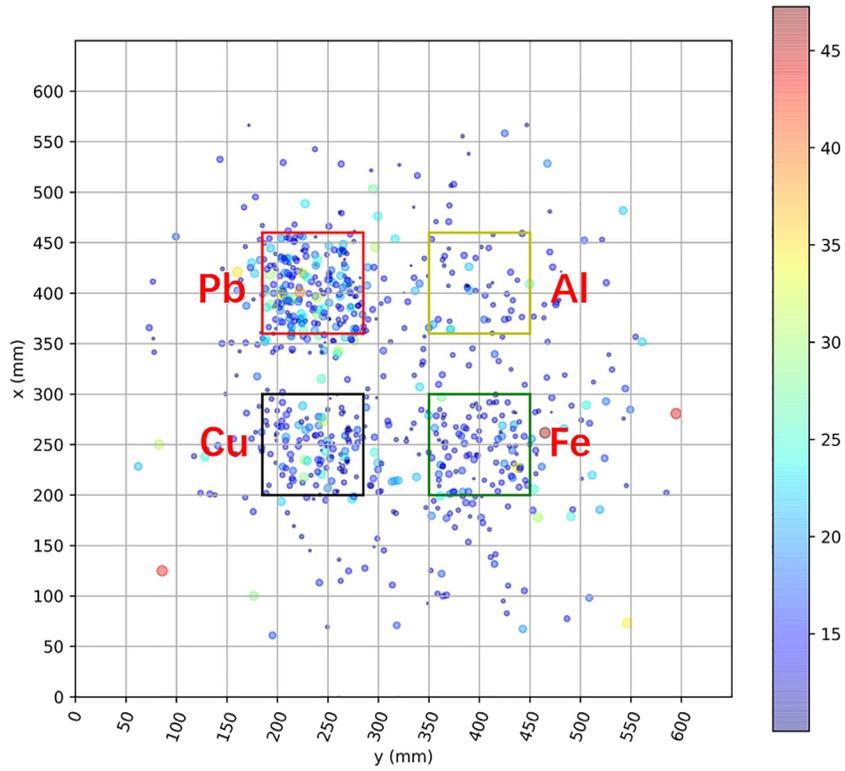

Figure 21: Image reconstruction of different materials

## 6.2 Material inspection alarm

A considerable amount of muons is necessary to get a clear image, but the flux of CRM is finite, this requires a long measurement time. In some scenarios, long time measurements are not acceptable, such as customs rapid inspection. Therefore, the quick alarm function is desired for such situations. A statistical tool, the Scattering Density Estimation (SDE) method [34], was employed by Canadian researchers to analyze the short-term measurement results and quickly give the judgment result of the presence or absence of high-$Z$ substances [23].

The rapid alarm method studied in this work is based on the two sets of data: with and without a 10 cm×10 cm×10 cm lead brick in the middle of the target layer. Firstly, the scattering position is calculated by PoCA algorithm, then the scattering density in each 3D voxel were obtained. Presence of high $Z$ object can be judged by comparing maximal value of the scattering density of all voxels with a preset SDE value. Based on the slices of two sets experimental data by time as shown in Figure 22, we could get the preset threshold (the value of intersection of the red dashed line and the horizontal axis). The preset threshold determines the accuracy of judgment, in order to analyze the confidence of the judgment result, we sliced the data with and without lead bricks according to 120 s, 360 s, and 720 s, counting the maximum scattering density of each voxel, and the distribution is shown in Figure 22. Each data filled into the histogram represents a data slice (120 s, 360 s and 720 s). The blue line represents the data without the lead bricks, and the orange line indicates data with lead bricks. Data falling on the right side of the preset threshold means the judgment is positive, and falling on the left side means the judgment is negative. Obviously, under the premise of the lead brick presence, the part of the data exceeding the threshold is True-Positive case; under the condition without brick, the part of the data exceeding the threshold is False-Positive case. Herein, the Receiver Operating Characteristic (ROC) [35] curves corresponding to 120 s, 360 s, and 720 s were drawn, as shown in Figure 23

It can be seen that: as the measurement time increases (120 s→360 s→720 s), the angular distribution within the voxels gradually approaches the true distribution and the obtained scattering density values are closer to the true value of the material. As shown in Figure 22 and Figure 23, respectively, the scattering density distribution of lead bricks presence data shifting to the right and the increasing of the area under the ROC curve (0.7861→ 0.8734→ 0.9211) with the increase of the length of the time slice.

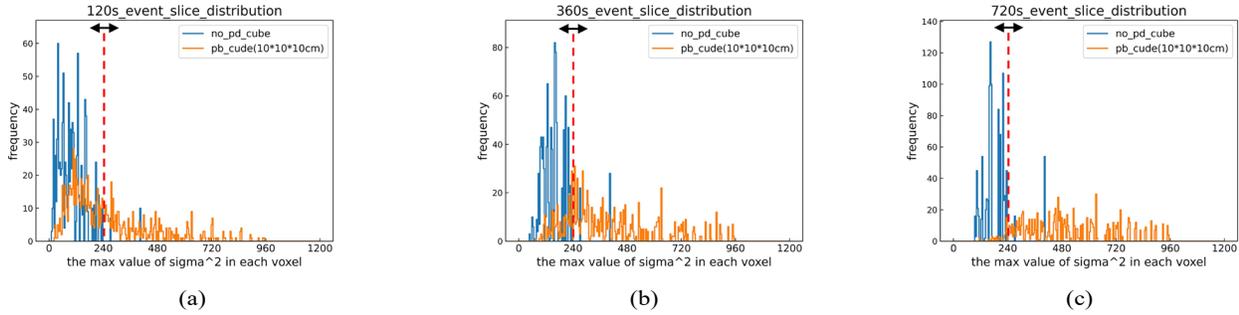

Figure 22: Scattering density of different length of time slices

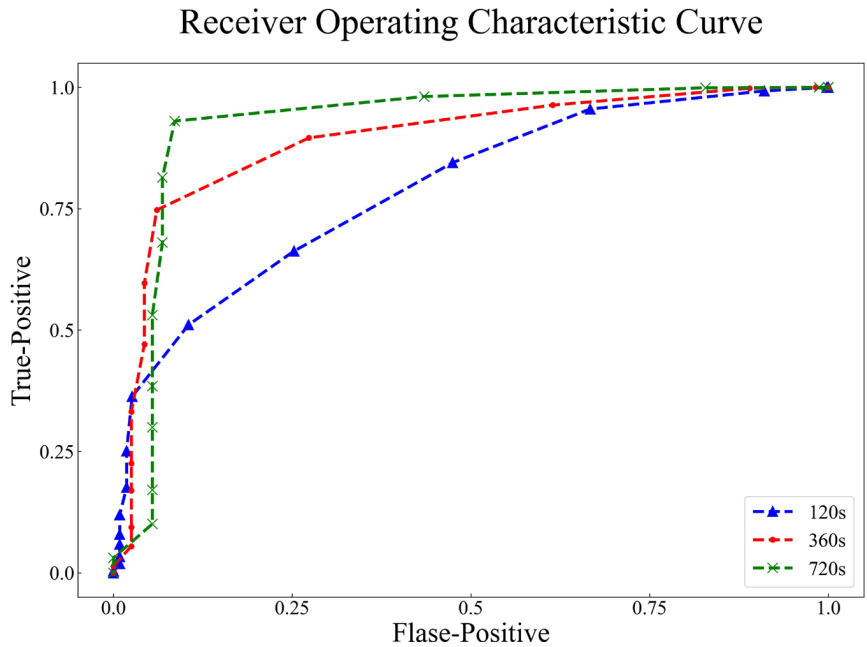

Figure 23: The Receiver Operating Characteristic curve

From Figure 22, it can be noticed that the SDE curve for lead-free bricks data(blue) has a peak at around 360, this peak is caused by the large-angle scattering events, the presence of this peak increases the probability of misjudgment when making a single determination. Therefore, we try to reduce this influence by making multiple determinations in a short period of time. Based on three times of 120 s measurements, if there are two or more times results show Positive, the system will alarm to report the existence of lead bricks, with the corresponding confidence level of 0.9049. Otherwise, it will be considered as without lead bricks (0.0951) case. The benefits of using multiple judgments are not only ensuring more than 90% of judgment accuracy but also limiting time consumption. The rapid alarm function expands the muography application on customs inspection.

## 7. Conclusion

In this work, a low-cost, compact and robust cosmic-ray muon prototype based on plastic scintillators was developed and commissioned. Its individual layer detection efficiency can reach about 98%. Meanwhile, it can achieve a position resolution of 2.50 mm and angular resolution of 8.73 mrad with the separation distance of 40.5 cm. Moreover, we conducted imaging tests, on the one hand, the feasibility of the inversion algorithm via Monte Carlo simulation data was verified; on the other hand, we completed the three-dimensional imaging of lead bricks experimentally. Pb, Cu, Fe and Al were also be imaged, and these metals could be clearly identified from the imaging results. In order to realize the rapid alarm function for high-$Z$ materials, the SDE method was used to identify the lead bricks, and ROC curve was used to evaluate the results. Enlarging the single sliced data time within the time allowable range can improve the accuracy of the adjustments. LUMIS has completed various tests, its feasibility in customs inspection has been verified. These lay the foundation of technology and craft to construct low cost, realistic, compact imaging devices that can be used in many on-site application scenarios.

## 8. Acknowledgement

The authors would like to acknowledge the support of the National natural Science Foundation of China (11975115), Special Projects of the Central Government in Guidance of Local Science and Technology Development (Research and development of three-dimensional prospecting technology based on Cosmic-ray muons, YDZX20216200001297), the Research and Development of Medical Isotopes based on High-current Superconducting Linear Accelerator Project, the Fundamental Research Funds for the Central Universities (lzujbky-2019-54) and the Science and Technology Planning Project of Gansu (20JR10RA645).

## References

[1] M. Tanabashi, K. Hagiwara, K. Hikasa, K. Nakamura, Y. Sumino, F. Takahashi, J. Tanaka, K. Agashe, G. Aielli, C. Amsler, et al., Review of particle physics, Physical Review D 98 (2018) 030001.

[2] V. Makhmutov, Y. I. Stozhkov, G. Bazilevskaya, N. Svirzhevsky, A. Morzabayev, Lati- tude effect of muons in the earth's atmosphere during solar activity minimum, Bulletin of the Russian Academy of Sciences: Physics 73 (2009) 350–352.

[3] M. Storini, M. Laurenza, Solar activity effects on muon data, MEMORIE-SOCIETA

ASTRONOMICA ITALIANA 74 (2003) 774–777.

[4] C. Augusto, C. Navia, M. Robba, Search for muon enhancement at sea level from transient solar activity, Physical Review D 71 (2005) 103011.

[5] A. Y. Konovalova, I. Astapov, N. Barbashina, N. Osetrova, Y. Mishutina, V. Shutenko,

I. Yashin, Analysis of muon flux variations caused by high-speed solar wind during periods of low solar activity, Physics of Atomic Nuclei 82 (2019) 909–915.

[6] X.-W. MENG, Effective temperature calculation and monte carlo simulation of tem- perature effect on muon flux, Chinese Physics C 28 (2004) 110–115.

[7] A. Dmitrieva, R. Kokoulin, A. Petrukhin, D. Timashkov, Corrections for temperature effect for ground-based muon hodoscopes, Astroparticle Physics 34 (2011) 401–411.

[8] A. Dmitrieva, I. Astapov, A. Kovylyaeva, D. Pankova, Temperature effect correction for muon flux at the earth surface: estimation of the accuracy of different methods, in: Journal of physics: Conference series, volume 409, IOP Publishing, p. 012130.

[9] P. K. Grieder, Cosmic rays at Earth, Elsevier, 2001.

[10] P. Checchia, Review of possible applications of cosmic muon tomography, Journal of Instrumentation 11 (2016) C12072.

[11] S. Procureur, Muon imaging: Principles, technologies and applications, Nuclear In- struments and Methods in Physics Research Section A: Accelerators, Spectrometers, Detectors and Associated Equipment 878 (2018) 169–179.

[12] L. Bonechi, R. D'Alessandro, A. Giammanco, Atmospheric muons as an imaging tool, Reviews in Physics 5 (2020) 100038.

[13] G. Bonomi, P. Checchia, M. D'Errico, D. Pagano, G. Saracino, Applications of cosmic- ray muons, Progress in Particle and Nuclear Physics 112 (2020) 103768.

[14] E. George, Cosmic rays measure overburden of tunnel, Commonwealth Engineer 455 (1955).

[15] L. W. Alvarez, J. A. Anderson, F. El Bedwei, J. Burkhard, A. Fakhry, A. Girgis,

A. Goneid, F. Hassan, D. Iverson, G. Lynch, et al., Search for hidden chambers in the pyramids, Science 167 (1970) 832–839.

[16]  K. Nagamine, M. Iwasaki, K. Shimomura, K. Ishida, Method of probing inner-structure of geophysical substance with the horizontal cosmic-ray muons and possible application to volcanic eruption prediction, Nuclear Instruments and Methods in Physics Research Section A: Accelerators, Spectrometers, Detectors and Associated Equipment 356 (1995) 585–595.

[17]  H. Tanaka, K. Nagamine, N. Kawamura, S. Nakamura, K. Ishida, K. Shimomura, Development of a two-fold segmented detection system for near horizontally cosmic-ray muons to probe the internal structure of a volcano, Nuclear Instruments and Methods in Physics Research Section A: Accelerators, Spectrometers, Detectors and Associated Equipment 507 (2003) 657–669.

[18]  K. Morishima, M. Kuno, A. Nishio, N. Kitagawa, Y. Manabe, M. Moto, F. Takasaki, H. Fujii, K. Satoh, H. Kodama, et al., Discovery of a big void in khufu's pyramid by observation of cosmic-ray muons, Nature 552 (2017) 386–390.

[19]  R. Nishiyama, A. Ariga, T. Ariga, A. Lechmann, D. Mair, C. Pistillo, P. Scampoli, P. Valla, M. Vladymyrov, A. Ereditato, et al., Bedrock sculpting under an active alpine glacier revealed from cosmic-ray muon radiography, Scientific reports 9 (2019) 1–11.

[20]  K. N. Borozdin, G. E. Hogan, C. Morris, W. C. Priedhorsky, A. Saunders, L. J. Schultz, M. E. Teasdale, Radiographic imaging with cosmic-ray muons, Nature 422 (2003) 277–277.

[21]  K. Gnanvo, L. V. Grasso, M. Hohlmann, J. B. Locke, A. Quintero, D. Mitra, Imaging of high-z material for nuclear contraband detection with a minimal prototype of a muon tomography station based on gem detectors, Nuclear Instruments and Methods in Physics Research Section A: Accelerators, Spectrometers, Detectors and Associated Equipment 652 (2011) 16–20. Symposium on Radiation Measurements and Applications (SORMA) XII 2010.

[22]  J. Cheng, X. Wang, Z. Zeng, The research and development of a prototype cosmic ray muon tomography facility with large area mrpc detectors, Nuclear Electronics and Detection Technology 34 (2014) 613–617.

[23]  V. Anghel, J. Armitage, F. Baig, K. Boniface, K. Boudjemline, J. Bueno, E. Charles, P. Drouin, A. Erlandson, G. Gallant, et al., A plastic scintillator-based muon tomogra- phy system with an integrated muon spectrometer, Nuclear Instruments and Methods

in Physics Research Section A: Accelerators, Spectrometers, Detectors and Associated Equipment 798 (2015) 12–23.

[24]   D. Mahon, A. Clarkson, D. Hamilton, M. Hoek, D. Ireland, J. Johnstone, R. Kaiser, T. Keri, S. Lumsden, B. McKinnon, et al., A prototype scintillating-fibre tracker for the cosmic-ray muon tomography of legacy nuclear waste containers, Nuclear Instruments and Methods in Physics Research Section A: Accelerators, Spectrometers, Detectors and Associated Equipment 732 (2013) 408–411.

[25]   C. Liao, H. Yang, Z. Liu, J. P. Hayward, Design and characterization of a scintillator- based position-sensitive detector for muon imaging, Nuclear Technology 205 (2019) 736–747.

[26]   Z. Liang, T. Hu, X. Li, Y. Wu, C. Li, Z. Tang, A cosmic ray imaging system based on plastic scintillator detector with sipm readout, Journal of Instrumentation 15 (2020) C07033.

[27]   SP101, http://www.hoton.com.cn/item/624739/189245/, 2022.

[28]   Si-PM, https://www.onsemi.com/pdf/datasheet/microc-series-d.pdf, 2022.

[29]   H. Liu, Z. Shen, Z. Liu, C. Feng, S. Liu, Development of the readout system for sipm-scintillator-based muography devices, Journal of Instrumentation 16 (2021) T06012.

[30]   Python Software Foundation, Pyqt5, 2021.

[31]   C. Hagmann, D. Lange, D. Wright, Cosmic-ray shower generator (cry) for monte carlo transport codes, in: 2007 IEEE Nuclear Science Symposium Conference Record, vol- ume 2, pp. 1143–1146.

[32]   D. Sunday, Practical geometry algorithms: With c++ code, 2021.

[33]   G. R. Lynch, O. I. Dahl, Approximations to multiple coulomb scattering, Nuclear Instruments and Methods in Physics Research Section B: Beam Interactions with Ma- terials and Atoms 58 (1991) 6–10.

[34]   V. Anghel, G. Jonkmans, C. Jewett, M. Thompson, Detecting high atomic number materials with cosmic ray muon tomography, 2015. US Patent 9,035,236.

[35]   T. Fawcett, An introduction to roc analysis, Pattern Recognition Letters 27 (2006) 861–874. ROC Analysis in Pattern Recognition.